\documentclass{article}
\usepackage{epsfig}

\begin{document}
\voffset 1.5cm

\begin{center}

{\LARGE \bf Electromagnetic properties in $^{160-170}$Dy nuclei: A microscopic description by the pseudo-SU(3) shell model}

\vspace{0.5cm}
Carlos E. Vargas$^{a}$\footnote{corresponding author; cavargas@uv.mx}, V\'ictor Vel\'azquez$^{b}$, 
Sergio Lerma-Hern\'andez$^{a, c}$ \footnote{On sabbatical leave from Facultad de F\'isica, Universidad Veracruzana}, Norma Bagatella-Flores$^{a}$

\vspace{0.5cm}
$^{a}$ Facultad de F\'isica, Universidad Veracruzana, CP 91000 Xalapa, Veracruz, M\'exico \\

$^{b}$Facultad de Ciencias, Universidad Nacional Aut\'onoma de M\'exico, Apartado Postal 70-542, 04510 M\'exico D.F., M\'exico \\

$^{c}$ Instituto de Ciencias Nucleares, Universidad Nacional Aut\'onoma de M\'exico -
Apdo. Postal 70-543, M\'exico D. F., C.P. 04510\\
\end{center}

\vskip 0.5cm
\date{\today}

\begin{abstract}

The large collectivity observed in the rare earth region of the nuclear landscape is well known.
The microscopic studies are difficult to perform in this region due to the enormous 
size of the valence spaces, a problem that can be avoided by means of the use of symmetry based models.
Here we present calculations for electromagnetic properties of $^{160-170}$Dy nuclei within the pseudo-SU(3) scheme.
The model Hamiltonian includes the preserving symmetry $Q\cdot Q$ term and the 
symmetry-breaking Nilsson and pairing terms, systematically parametrized for all members of the chain. 
The model is used to calculate B(E2) and B(M1) inter-band transition strengths between the ground state, 
$\gamma$ and $\beta$-bands. In addition, we present results for quadrupole moments and $g$ factors 
in these rotational bands. The results show that the pseudo-SU(3) shell model is a powerful microscopic
theory for a description of electromagnetic properties of states in the normal parity sector in 
heavy deformed nuclei.

\end{abstract}

\section{Introduction}
Electromagnetic properties of a nuclear system can provide valuable information on the microscopic structure, because they are very sensitive to the single-particle aspects of the nuclear wave function and hence can serve as a strict testing ground for theoretical models. Among these, the g-factors may depend on the microscopic structure, whereas those related to the electric quadrupole transitions allow us to determine the overall collectivity of the system. In relation with the experiments, the existence of unfavorable kinematic matching conditions leading to heavy neutron-rich nuclei makes the measurements difficult, because the small cross sections pose limitations to its use for populating these nuclei \cite{Fir96,Wu98}. In fact, even with current high-intensity fragmentation facilities, the frontier of measured neutron-rich dysprosium nuclei does not reach beyond the $^{174-176}$Dy isotopes \cite{Sod16,Wat16}, and the greatest efforts are being made in the measurement of K-isomers for the Dysprosium isotopes. For stable isotopes, low-excitation B(E2), quadrupole moments and $g_{\gamma}/g_{g.s.}$ ratios are good candidates for future experiments, which makes interesting the predictions of these observables. 

Regarding the use of theoretical schemes, the large valence spaces associated with heavy-nuclei have implied a slow progress, limiting the number of microscopic models capable of dealing with rare-earth nuclei. A shell model description of heavy nuclei requires assumptions that include a systematic and proper truncation of the Hilbert space. The symmetry based SU(3) shell model \cite{Ell58,Ell58b} has been successfully applied in light nuclei (A$<$50), where a harmonic oscillator mean-field and a residual quadrupole-quadrupole interaction can be used to describe dominant features of the nuclear spectra. On the other hand, this model becomes useless in heavy nuclei due to the breaking of the SU(3) symmetry by the spin-orbit coupling, while at the same time pseudospin emerges as a good symmetry \cite{Hec69,Ari69}. This new symmetry has its origin in the relativistic mean field for heavy-nuclei \cite{Gin97,Blo95}, and its consistency has determined the success of the pseudo-SU(3) model \cite{Rat73}. It refers to the well known quasidegeneracy observed in heavy nuclei between single-nucleon orbitals with $j=l-1/2$ and $j=(l-2)+1/2$ in the shell $\eta$. These orbitals can therefore be labeled as pseudospin doublets with quantum numbers $\tilde{j}=j$, $\tilde{\eta}=\eta-1$, and $\tilde{l}=l-1$. A different approach that includes a systematic truncation scheme is the Projected Shell Model. This is a model where the projection techniques is an efficient way of truncating the shell model space, which is otherwise too large to handle. With that model has been possible to describe, among others, excitated states in $^{160,162}$Dy nuclei \cite{Jun02} or B(E2) values for inter-band transitions between the g.s., $\gamma$-, and $\gamma \gamma$-bands in $^{166,168}$Er \cite{Bou02}.

More than four decades have passed since the pseudo-SU(3) model was first introduced. Initially the pseudo-spin was considered as a dynamical symmetry \cite{Cas87}. With the development of computer code it has been possible to calculate reduced matrix elements of physical operators (i.e. the breaking-symmetry Nilsson single-particle energies or pairing correlations) between different pseudo-SU(3) irreps \cite{Bah94}. More recently, especially over the last few years, this enabling technology has been used to describe a wide variety of very different nuclear phenomena. Among others applications, it allows fully microscopic studies of energy levels belonging to normal parity bands \cite{Var00}, B(E2) intra- and inter-band transitions \cite{Var01}, the high density of $0^+$ states in $^{158}$Gd \cite{Hir06} and the analysis of the interplay between orbital and spin components of the scissors mode \cite{Var03} in both even-even and A-odd nuclei. In a recent work, a systematic parametrization of the interaction in the frame of the pseudo-SU(3) model \cite{Var13} for the $^{160-170}$Dy chain of dysprosium isotopes has allowed for the calculation of energies and intra-band B(E2) strengths in the ground state, $\gamma $ and $ \beta$-bands. Some preliminary results, concerning the electric quadrupole inter-band transitions have already been reported by some of us in Ref. \cite{Var15}.

In this Article, our goal is to study the electric quadrupole moment, B(E2) inter-band transition strengths,
g-factors and M1 transitions in ground state, $\gamma$ and $\beta$ bands along the dysprosium isotopic chain (Z=66) starting at N = 94 ($^{160}$Dy) and ending with the N = 104 midshell $^{170}$Dy nucleus. 
To this end, we have taken the results of the pseudo-SU(3) model 
of Reference \cite{Var13}, where the Nilsson, Quadrupole-Quadrupole and Pairing terms of the Hamiltonian 
have been systematically parametrized as a function of the mass ($A$) \cite{Rin79}, and a best fit to experimental data of the parameters $a(K^2)$, $b(J^2)$ and $c(\tilde{C}_3)$ has been found for the $^{160-170}$Dy isotopic chain. With those results at hand, we take the wave functions of each state in order to calculate the various electromagnetic properties in the considered chain.

This contribution is organized as follows: the group theory that underlies the pseudo-SU(3) 
model is reviewed in the first part of the next section. A very limited number of results 
are given because a rather extensive discussion of the correspondence between the quantum 
numbers, Casimir operators, and pseudo-SU(3) wave functions can be found in previously
published articles \cite{Cas87,Tro95,Tro95b}. The pseudo-SU(3) Hamiltonian and a choice for the 
corresponding interaction strength parameters are also reviewed \cite{Var13}, as they are of special importance for a theoretical description of the $^{160-170}$Dy isotopes. In the final part of Sect. \ref{model}, definitions of experimentally measured quantities are reviewed. A comparison of pseudo-SU(3) model results to the available experimental data for the electric quadrupole moments and B(E2) transition strengths are presented in Sect. \ref{electric}, and the results for the g-factors and M1 transitions are presented in Sect. \ref{magnetic}. A summary as well as a brief conclusion are given in Sect. \ref{summary}.

\section{The model}\label{model}

As described by Troltenier {\it et al.} in Ref. \cite{Tro95}
the pseudo-SU(3) model is a many-particle shell-model based theory that takes full advantage 
of pseudospin symmetry, which, in heavy nuclei, is manifest in the near degeneracy of the orbital 
pairs $[(l-1)_{j=l-1/2},(l+1)_{j=l+1/2}]$. It is also a theory that takes full account of
the Pauli Exclusion principle. Like most other shell-model schemes, the proton and neutron 
configuration spaces of the pseudo-SU(3) model are usually restricted to a single oscillator shell. 
A recent extension of the pseudo-SU(3) model takes the $\tilde{S}_{\pi,\nu} = 0$ and $1$ (subscript $\pi$ for proton and $\nu$ for neutron) spin degrees of freedom into account in a full proton-neutron formalism \cite{Var04}. The scheme 
is an algebraic shell-model theory that exploits powerful group theoretical 
methods in the construction of basis functions and also the calculation of required matrix elements. 
Specifically, basis states are labeled by eigenvalues of Casimir operators of the underlying
symmetry groups and additional indices that are required to resolve multiplicities in the group 
reductions. Since a full discussion of these matters can be found elsewhere \cite{Cas87}, 
only the results will be included in this contribution. In terms of the space $(U(N) \leftrightarrow 
[f])$, shape $(SU(3) \leftrightarrow (\lambda, \mu))$, orbital $(SO(3) \leftrightarrow L)$, spin $(S)$
and total angular momentum $(J)$ as well as the various multiplicities ($\alpha$ for the $U(N) \supset 
SU(3)$ reduction, $\rho$ for the product of two SU(3) irreps, and $\kappa$ for the $SU(3) \supset SO(3)$
reduction) the basis states have the form:

\begin{eqnarray}
| \{ \alpha_\pi (\lambda_\pi, \mu_\pi)\kappa_\pi S_\pi, \alpha_\nu (\lambda_\nu, 
\mu_\nu)\kappa_\nu S_\nu  \} \rho (\lambda, \mu) \kappa L S; JM \rangle.
\end{eqnarray}

The occupation numbers for protons are constant along the chain. In shell model applications, 
the dysprosium isotopes are considered to have 16 protons out of the Z = 50 inert core, 
10 of these in normal and 6 in abnormal $h_{11/2}$ parity levels. For neutrons, the occupation 
numbers in normal parity levels change from 8 to 14. What makes the pseudo-SU(3) model a powerful 
theory is that it allows one to invoke a relatively simple and physically motivated basis truncation 
scheme. Note, however, that the configuration space that was used is composed of states with 
zero and one proton and neutron pseudospin. This corresponds to selecting U(N) irreps with spin zero
and one with the highest spatial symmetry, which have a non negligible contribution to excited 
rotational bands. The virtue of the pseudo-SU(3) symmetry is preserved by taking care that states with
$\tilde{S}_{\pi,\nu}=0$ should be dominant in the ground state. It translates into severe limits 
for the ``rotorlike" terms, and guarantees that the whole band structure is preserved.
The effects resulting from higher spin excitations (beyond $\tilde{S}_{\pi,\nu}=1$) are suppressed, 
which is not a too crude approximation because higher spin-flip modes have been determined by 
experiment to lie at significantly higher energies than most of the states considered in the present 
analysis.

The Hamiltonian contains spherical Nilsson single-particle terms for the protons 
and neutrons ($H_{sp,\pi[\nu]}$), the quadrupole-quadrupole ($\tilde Q \cdot \tilde 
Q$) and pairing ($H_{pair,\pi[\nu]}$) interactions parametrized systematically, as 
well as three rotorlike terms ($K^2$, $J^2$ and $\tilde C_3$) that are diagonal in the SU(3) basis:

\begin{eqnarray}
    H & = & \sum_{\alpha=\pi,\nu} \{ H_{sp,\alpha} - G_\alpha ~H_{pair,\alpha}
	\} - \frac{1}{2}~  \chi~ \tilde Q \cdot \tilde Q \label{eq:ham} \\
      &   & + ~a~ K^2 + ~b~ J^2~ + ~c~ \tilde C_3. \nonumber
\end{eqnarray}

\noindent A detailed analysis of each term of this Hamiltonian and its parametrization can be 
found in Ref. \cite{Var00b}. The first row contains the basic components of any realistic 
Hamiltonian: the single-particle levels, pairing correlations and the quadrupole-quadrupole 
interaction, essential in the description of deformed nuclei. They have been widely studied 
in nuclear physics literature, allowing one to fix their respective strengths by systematics.
The rotorlike terms in the Hamiltonian (eq. \ref{eq:ham}) are used to fine tune the energies in ground-state, 
$\gamma$ and $\beta$ bands. Their three parameters $a$ (= 20 keV), $b$ (= -3.2 keV), and $c$ (= 0.033 keV) were determined by applying a best fit to the experimental data in $^{160-168}$Dy. Further details on the procedure followed to fit them can be found in Ref \cite{Var13}.

Before discussing comparisons of electromagnetic properties to experimental data, it is 
important to write down definitions for transition probabilities and the electromagnetic moments.

The magnetic dipole transition operator, following the notation of Ref. \cite{Cas87,Tro95b}, is given by
\begin{eqnarray}
T_{1\mu}(M1) \equiv \mu_N \left( g_\pi^o L_{1\mu}^\pi + g_\nu^o L_{1\mu}^\nu 
+  g_\pi^s S_{1\mu}^\pi + g_\nu^s S_{1\mu}^\nu \right) \label{mdo}
\end{eqnarray}
where $L_\mu^\sigma$ ($S_\mu^\sigma$) and $g^o_\sigma$ ($g^s_\sigma$) are the orbital (spin) angular momentum and orbital (spin) gyro-magnetic factor, respectively, for protons ($\sigma = \pi$) and neutrons
($\sigma = \nu$), and $\mu_N$ denotes the nuclear magneton. In the present work, the orbital and 
‘quenched’ (by a factor of 0.7) spin $g$ factors for protons and neutrons are used: $g_\pi^o = 1$,
$g_\nu^o = 0$, $g_\pi^s = (0.7)5.5857$ and $g_\nu^s = -(0.7)3.8263$.

Similarly, the electric quadrupole transition 
operator is defined as
\begin{eqnarray}
T_{2\mu}(E2) & \equiv & b_0^2 \left( e_\pi \sum_{i_\pi} \sqrt{\frac{16\pi}{5}} r_\pi^2(i) 
Y_{2\mu}(\theta_i^\pi,\phi_i^\pi) \right) \nonumber \\
&  & + b_0^2 \left( e_\nu \sum_{i_\nu} \sqrt{\frac{16\pi}{5}} r_\nu^2(i) 
Y_{2\mu}(\theta_i^\nu,\phi_i^\nu) \right)\nonumber.
\end{eqnarray}

The SO(3) reduced matrix elements of a tensor operator $T_{JM}$ between states of 
initial (final) angular momentum and projection $J_i$ and $M_i$ ($J_f$ and $M_f$) is 
defined \cite{Tro96} by 
\begin{eqnarray}
\langle \gamma_f J_f || T_J || \gamma_i J_i \rangle \langle J_iM_i , JM | J_fM_f \rangle 
\equiv \langle \gamma_f J_f M_f | T _{JM} | \gamma_i J_i M_i \rangle \nonumber
\end{eqnarray}
where $\gamma_i$ and $\gamma_f$ represent additional quantum numbers, and $\langle \gamma_f J_f || T_J || 
\gamma_i J_i \rangle$ is the reduced matrix element. From this definition, it follows 
that the reduced probability for electric quadrupole transition \cite{Tro95} is given by 
\begin{eqnarray}
B(E2;\gamma_i J_i \rightarrow \gamma_f J_f) \equiv \frac{2J_f+1}{2J_i+1}
\langle \gamma_f J_f || Q_2 || \gamma_i J_i \rangle^2 \nonumber
\end{eqnarray}
and that for magnetic dipole radiation by
\begin{eqnarray}
B(M1;\gamma_i J_i \rightarrow \gamma_f J_f) \equiv \frac{2J_f+1}{2J_i+1}
\langle \gamma_f J_f || M_1 || \gamma_i J_i \rangle^2 \nonumber.
\end{eqnarray}

The definition of the electric quadrupole moment is given by
\begin{eqnarray}
Q(\gamma J) \equiv \sqrt{\frac{16\pi}{5}} \sqrt{\frac{J(2J-1)}{(J+1)(2J+3)}}
\langle \gamma J || Q_2 || \gamma J \rangle \nonumber
\end{eqnarray}
and that of the magnetic dipole moment, following the notation of Ref. \cite{Cas87,Tro96}, by
\begin{eqnarray}
\mu(\gamma J) = {\langle \gamma JM=J |M_{10}| \gamma JM=J \rangle}/{J}
\end{eqnarray}
from which the definition of the gyro-magnetic factor is determined to be
\begin{eqnarray}
g(\gamma J)=\frac{\mu(\gamma J)}{J}.
\end{eqnarray}

In the pseudo-SU(3) dynamical symmetry limit, no transitions between ground state band and the $\beta$-band are predicted. The reason for this behaviour relies on the fact that the ground state band and the lowest excited $K^\pi=0^+$ band belong to different representations of the SU(3) group. The lowest excited $K^\pi=0^+$ band has the labels ($\lambda=2N-4,\mu=2$) whereas the ground state band belongs to the ($\lambda=2N,\mu=0$) representation \cite{Leh99}. As the E2 transition operator consists of generators of the SU(3) group, no transitions will be induced between different representations. Therefore, no transitions between the ground state and the lowest-excited $K^\pi=0^+$ band are predicted in the SU(3) dynamical limit. In contrast, if one disturbs the dynamical pseudo-SU(3) symmetry, these transitions are allowed, but their values are far lower than the transitions within the band.

\section{B(E2) strengths and quadrupole moments}\label{electric}

As indicated above, in this contribution the pseudo-SU(3) model is used to investigate 
electromagnetic properties of $^{160-170}$Dy. In Table 
\ref{be2-inter}, the results for inter-band B(E2) strengths between states of 
ground state, $\gamma$ and $\beta$-bands are presented. These values are significantly smaller than the
intra-band B(E2) strengths discussed in Table 2 of Ref. \cite{Var13}, because the
wave functions of states belonging to different bands have components in almost completely 
different $SU(3)$ irreps. Nevertheless, 
we find some inter-band B(E2) strengths with large values, which gives evidence of a strong overlap 
between the wave functions of the states due to the breaking symmetry terms in the Hamiltonian
(\ref{eq:ham}).

\begin{table}
\begin{tabular}{c|cc|cc|cc|c|c|c}\hline \hline
 & \multicolumn{9}{c}{B(E2) [$e^2b^2 \times 10^{-2}$]} \\
$J_{i,band}^{\pi} \rightarrow J_{f,band}^{\pi}$ & \multicolumn{2}{c}{$^{160}$Dy} & \multicolumn{2}{c}{$^{162}$Dy} & \multicolumn{2}{c}{$^{164}$Dy} & $^{166}$Dy & $^{168}$Dy & $^{170}$Dy \\ 
                                                & Exp. & Theo.& Exp. &Theo. & Exp. &Theo. & Theo.& Theo.& Theo. \\ \hline
$0^+_{gsb}  \rightarrow  2^+_{\gamma}$  & 11.6$\pm$0.8 & 14.4 &12.1$\pm$0.8&14.6&10.7$\pm$1.1&16.5& 2.3 & 1.7   & 18.4 \\
$2^+_{gsb}  \rightarrow  4^+_{\gamma}$          &      &  1.6 &      &  1.6     &  & 0.7     & 0.4      & 0.4   & 0.5  \\
$4^+_{gsb}  \rightarrow  5^+_{\gamma}$          &      &  4.0 &      &  4.1     &  & 3.4     &  0       &  0    & 4.7  \\
$4^+_{gsb}  \rightarrow  6^+_{\gamma}$          &      &  0.2 &      &  0.2     &  & 0.1     & 0.1      & 0.1   & 0.1  \\
$6^+_{gsb}  \rightarrow  7^+_{\gamma}$          &      &  2.1 &      &  2.2     &  & 1.1     &  0       &  0    & 2.6  \\
$6^+_{gsb}  \rightarrow  8^+_{\gamma}$          &      &   0  &      &  0.2     &  & 1.2     &  0       &  0    & 0.3  \\
$8^+_{gsb}  \rightarrow  9^+_{\gamma}$          &      &  0.6 &      &  0.7     &  & 0.1     & 0.1      & 0.1   & 1.3  \\
$8^+_{gsb}  \rightarrow 10^+_{\gamma}$          &      &   0  &      &  0.9     &  & 2.5     &  0       & 0.1   & 1.0  \\
$0^+_{gsb}  \rightarrow  2^+_{\beta}$   &  1.7$\pm$0.2 &  0.2 &      &   0      &  &  0      & 1.0      & 1.1   & 0.4  \\
$2^+_{gsb}  \rightarrow  4^+_{\beta}$           &      &  0.3 &      &   0      &  &  0      & 0.4      & 0.4   & 0.1  \\
$4^+_{gsb}  \rightarrow  6^+_{\beta}$           &      &  0.1 &      &   0      &  & 0.1     &  0       &  0    &  0   \\
$6^+_{gsb}  \rightarrow  8^+_{\beta}$           &      &   0  &      &   0      &  & 0.1     & 0.2      & 0.2   &  0   \\
$0^+_{\beta}  \rightarrow  2^+_{\gamma}$        &      &  4.4 &      &  4.4     &  & 3.4     & 180.5    & 173.6 &  0.4 \\
$2^+_{\beta}  \rightarrow  3^+_{\gamma}$        &      &  1.4 &      &  1.1     &  & 1.8     & 108.4    & 113.5 & 19.1 \\
$2^+_{\beta}  \rightarrow  4^+_{\gamma}$        &      &  0.4 &      &  0.3     &  & 1.7     &  26.6    &  27.7 & 11.5 \\
$4^+_{\beta}  \rightarrow  5^+_{\gamma}$        &      &  0.3 &      &  0.3     &  & 2.0     & 1.1      & 1.0   & 27.0 \\
$4^+_{\beta}  \rightarrow  6^+_{\gamma}$        &      &  0.2 &      &   0      &  & 1.7     &  0       &  0    & 33.6 \\
$6^+_{\beta}  \rightarrow  7^+_{\gamma}$        &      &   0  &      &  0.7     &  & 1.6     & 1.4      & 1.1   & 35.2 \\
$6^+_{\beta}  \rightarrow  8^+_{\gamma}$        &      &   0  &      &   0      &  & 1.6     & 1.1      & 0.4   & 57.1 \\
$8^+_{\beta}  \rightarrow  9^+_{\gamma}$        &      &  0.1 &      &  1.2     &  & 0.8     & 1.9      & 3.1   & 43.6 \\ \hline \hline
\end{tabular}
\caption{B(E2;$J^+_i \rightarrow J^+_f$) inter-band transition strengths in $^{160-170}$Dy nuclei [given in $e^2b^2 \times 10^{-2}$]. The first column gives the initial ($J_i$) and final ($J_f$) values of the angular momentum. From second to the tenth column are the experimental (Exp.) and pseudo-SU(3) model calculations (Theo.). Effective charges are $e_\pi$=2.3 and $e_\nu$=1.3.}
\label{be2-inter}\end{table}

Configuration space of the pseudo-SU(3) model is restricted to a single shell, meaning 
that core excitations and higher shell correlations have not been taken into account, therefore
polarization effects are not included. This is a common defect in shell-model theories and is 
usually compensated by the use of effective charges in the calculation of B(E2) values and quadrupole moments. The effective charges used in the electric quadrupole transition operator $Q_\mu$ are 
$e_\pi = 2.3$ and $e_\nu=1.3$. These values are the same used in the pseudo-SU(3) studies up to 
now allowing one to describe both intra- and inter-band B(E2)s. They are larger than those used in 
standard calculations of quadrupole transitions \cite{Rin79} due to the absence of nucleons 
in intruder levels, and they were not varied to fit any particular value. 

Some results listed in Table \ref{be2-inter} may be the starting point for a systematic analysis of the $0^+_2$ excitations and determine whether they are $\beta$ oscillations. As it has been pointed out \cite{Gar01}, the $B(E2)$ strengths between $\beta$ and ground state band display a wide variation of values over well deformed nuclei, with variations of orders of magnitude over a single isotopic chain. That means that we cannot continue our practice of labelling the $0^+_2$ state by $\beta$, irrespective of its properties. It is neccesary to evaluate the electric quadrupole transitions to determine whether the excited $0^+_2$ actually corresponds to a $\beta$ vibration. For example,  if the $B(E2; 2^+_{\gamma} \rightarrow 0^+_{gs})$ value is 5 Wu (a typical value in the deformed rare-earth region), the $B(E2; 0^+_{\beta} \rightarrow 2^+_{gs})$ value should be of the order of 12 Wu. In analogous manner, the $B(E2; 2^+_{\beta} \rightarrow 0^+_{gs})$ values should be of the order of 2.5 Wu for the band to be considered as a $\beta$-vibrational candidate. However, such work is beyond the scope of the present paper.

The predictions of quadrupole moments that were calculated within the framework of the pseudo-SU(3) model are listed in Table \ref{q-moments}.
\begin{table}
\begin{tabular}{c|cc|c|cc|c|c|c}\hline \hline
 & \multicolumn{8}{c}{$Q(J^\pi_{band})$ [eb]} \\
$J_{band}^{\pi}$ & \multicolumn{2}{c}{$^{160}$Dy} & $^{162}$Dy & \multicolumn{2}{c}{$^{164}$Dy} & $^{166}$Dy & $^{168}$Dy & $^{170}$Dy \\
                                                & Exp. & Theo.& Theo. & Exp. &Theo. & Theo.& Theo.& Theo. \\ \hline
$2^+_{gsb}$    &1.8$\pm$0.40 &-2.19& -2.20 &-2.08$\pm$0.15&-2.31&-2.42&-2.42& -2.40 \\
$2^+_{\gamma}$  &  &  2.16      &  2.16 & &  2.31 &  2.00 &  1.87 &  2.32 \\
$2^+_{\beta}$   &  &  0.15      &  0.03 & & -2.28 & -2.15 & -2.13 & -1.66 \\
$3^+_{\gamma}$  &  & -0.02      & -0.02 & &  0.01 & -0.03 & -0.05 & -0.02 \\
$4^+_{gsb}$     &  & -2.74      & -2.76 & & -2.84 & -3.03 & -3.04 & -3.00 \\
$4^+_{\gamma}$  &  & -1.20      & -1.21 & & -1.25 &  1.56 &  1.47 & -1.51 \\
$4^+_{\beta}$   &  & -1.49      & -1.37 & & -2.74 & -2.71 & -2.71 & -1.22 \\
$5^+_{\gamma}$  &  & -1.78      & -1.79 & & -1.82 &  0.82 &  0.87 & -1.93 \\
$6^+_{gsb}$     &  & -2.93      & -2.94 & & -2.95 & -3.25 & -3.26 & -3.21 \\
$6^+_{\gamma}$  &  & -2.38      & -2.42 & & -2.45 & -0.12 & -0.11 & -1.84 \\
$6^+_{\beta}$   &  & -1.92      & -2.07 & & -2.84 &  0.62 &  0.63 & -1.73 \\
\hline \hline
\end{tabular}
\caption{Quadrupole moments in $^{160-170}$Dy in units of [eb]. The column on the left gives the angular momentum $J$, band, and parity $\pi$; and the columns from second to ninth are the experimental (Exp.) and pseudo-SU(3) model calculations (Theo.) for $Q(J^\pi_\alpha)$.}
\label{q-moments}\end{table}
Note the agreement with the data in $^{160}$Dy and $^{164}$Dy. In the first case, the result of the model is within the experimental uncertainties, while in $^{164}$Dy it is only 0.08 [eb] out of range. The quadrupole term is diagonal in the SU(3) scheme, and therefore, as $Q \cdot Q$ begins to dominate the Hamiltonian, it not only drives the system towards larger $\beta$ and smaller (larger) $\gamma$ for shells which are less (more) than half filled, but it also sharpens the shape of the nucleus.

\section{G-factors and M1 transitions}\label{magnetic}

It is important to consider how the magnetic properties of $^{160-170}$Dy are treated in the 
pseudo-SU(3) model and for this it is necessary to recall the expression for the magnetic dipole 
operator in Eq. (\ref{mdo}). Since the configuration space for the current implementation of the
pseudo-SU(3) model is made up of states with $\tilde{S}_{\pi,\nu} = 0$ and $1$, the matrix elements of both
proton and neutron spin operators have a non negligible contribution. This implies a much more 
comprehensive treatment of magnetic properties, as in previous studies with the model \cite{Tro96}, 
where only the states of maximal space symmetry were considered, so that matrix elements of the 
proton and neutron spin operators in Eq. (\ref{mdo}) were null. Table \ref{g-factors} lists the model
predictions for $g$ values.

\begin{table}
\begin{tabular}{c|cc|cc|cc|c|c|c}\hline \hline
 & \multicolumn{9}{c}{g-factors $(J^\pi_{band})~[\mu_N]$} \\
$J_{band}^{\pi}$ & \multicolumn{2}{c}{$^{160}$Dy} & \multicolumn{2}{c}{$^{162}$Dy} & \multicolumn{2}{c}{$^{164}$Dy} & $^{166}$Dy & $^{168}$Dy & $^{170}$Dy \\
                 &     Exp.     & Theo. &   Exp.      & Theo. &    Exp.   & Theo. & Theo. & Theo. & Theo. \\ \hline
$2^+_{gsb}$      & 0.36$\pm$0.01& 0.35  & 0.34$\pm$0.01& 0.34 &0.34$\pm$0.01& 0.32&  0.26 &  0.26 &  0.34 \\
$2^+_{\gamma}$   & 0.40$\pm$0.03& 0.27  & 0.46$\pm$0.03& 0.27 &0.38$\pm$0.03& 0.30&  0.21 &  0.21 &  0.57 \\
$2^+_{\beta}$    &              & 0.29  &             & 0.24  &           & 0.32  &  0.31 &  0.30 &  0.40 \\
$3^+_{\gamma}$   &              & 0.27  &             & 0.28  &           & 0.30  &  0.29 &  0.28 &  0.47 \\
$4^+_{gsb}$      & 0.35$\pm$0.02& 0.35  & 0.28$\pm$0.03& 0.35 &0.25$\pm$0.03& 0.32&  0.27 &  0.27 &  0.34 \\
$4^+_{\gamma}$   &              & 0.28  &             & 0.28  &           & 0.29  &  0.23 &  0.23 &  0.36 \\
$4^+_{\beta}$    &              & 0.37  &             & 0.33  &           & 0.33  &  0.31 &  0.31 &  0.37 \\
$5^+_{\gamma}$   &              & 0.29  &             & 0.28  &           & 0.30  &  0.25 &  0.24 &  0.39 \\
$6^+_{gsb}$      & 0.35$\pm$0.02& 0.35  & 0.36$\pm$0.02& 0.35 &0.32$\pm$0.02& 0.32&  0.28 &  0.27 &  0.34 \\
$6^+_{\gamma}$   &              & 0.29  &             & 0.28  &           & 0.29  &  0.27 &  0.27 &  0.35 \\
$6^+_{\beta}$    &              & 0.36  &             & 0.35  &           & 0.35  &  0.25 &  0.25 &  0.36 \\
\hline \hline
\end{tabular}
\caption{Gyro-magnetic factors in $^{160-170}$Dy in units of the nuclear magneton, $[\mu_N]$. The column on the left gives the angular momentum $J$, parity $\pi$ and the corresponding band; and the columns from second to tenth list the experimental data \cite{Alf97,Bra99} and the predictions of the pseudo-SU(3) model for $g(J^\pi_{band})$.}
\label{g-factors}\end{table}

It is well known from observables such as $R_{42}$ that heavy nuclei make a rapid transition from spherical to deformed behavior, which is also evident in the g-factors of the rare-earth nuclei as well \cite{Stu12}. Large variations in the values of the $g(2^+_1)$ are usually found near the closed shells. For the N=82 nuclei, $g(2^+_1) \sim 1$, the specific value being determined by the particular proton configuration. Then, with the addition of two neutrons, $g(2^+_1)$ falls well below the collective trend. In particular, it has been found in $^{156-164}$Dy \cite{Bra99,Stu12} that experimental values of the $g(2^+_{gsb})$ factors are almost constant, finding a small decrease of $\sim$ 0.39 in $^{156}$Dy to $\sim$ 0.34 in $^{164}$Dy. The results of the pseudo-SU(3) model, presented in Table \ref{g-factors}, reproduce this trend, with a decrease in the magnitude of $g(2^+_1)$ from 0.35 in $^{160}$Dy to 0.32 in $^{164}$Dy. For $^{166}$Dy and $^{168}$Dy, the model predicts an even greater reduction to 0.26. This drop in the value of $g(2^+_1)$ to 0.26, however, is also found by the projected shell model \cite{Bia07,Sun16}.

An interesting relation between the ground state band and the $\gamma$-band g-factors has been sugested \cite{Alf96}, finding that the ratio $r=g(2^+_\gamma)/g(2^+_{gsb})$ could be important in determining nuclear triaxiality \cite{Sun02}. The triaxial projected shell model has been used to determine the value of $r$ in the isotopes of Dy, finding that when the number of neutrons approaches to the mid-shell (N $\sim$ 104), $r \sim 1$, which has been interpreted as a transition to the triaxiality. The values of the pseudo-SU(3) model show this transition towards triaxiality, however, the value of the ratio $r$ is not increased monotonously. For $^{160}$Dy and $^{162}$Dy, we find that $r \sim 0.78$ which is very close to 0.82$\pm$0.03 reported by \cite{Alf96} but in disagreement with the 1.14$\pm$0.09 reported by \cite{Bra99}; in $^{164}$Dy, $r \sim 0.94$; for $^{166}$Dy and $^{168}$Dy, $r \sim 0.81$ and for $^{170}$Dy, $r \sim 1.67$. A description of triaxial deformed many-particle configuration, which pairing favors \cite{Tro95}, implies a strong mixing of SU(3) basis states. While the quadrupole-quadrupole interaction preserves and enhances the SU(3) symmetry of nuclear systems, on the other side pairing force breaks SU(3). Only for mid-shell nuclei, when the Pauli Principle forces the most deformed irrep to be triaxial, the $Q \cdot Q$ and pairing forces actually favor the same triaxial distribution. A more detailed study of the microscopic origin of the differences in the values between $g(2^+_\gamma)$ and $g(2^+_{gsb})$ and its relation to the Hamiltonian employed is left for future work. 

In addition to the previous analysis, the evolution of the g-factors as a function of angular momentum has been studied. In particular, for the isotopes $^{158-162}$Dy \cite{Alf97} has been found that for $J=6$, the g-factors fall to its minimum value and increase again when $J=8$. This fall in value by increasing J has been explained by the rotational alignment of $i_{13/2}$ neutrons, which have a negative g-factor. Nevertheless, the theoretical study using Hartree-Fock-Bogoliubov cranking wave functions \cite{Die80}, gave the result that $g(I)$ should decrease quite rapidly with rising spin for $^{156}$Dy, somewhat less rapidly for $^{158}$Dy and it should be nearly constant for $^{164}$Dy. A similar conclusion has been obtained with angular-momentum-projected Tamm-Dancoff theory \cite{Sun94}, the PSM \cite{Vel99} as well as with other fenomenological models \cite{Zha06} for nuclei toward the midshell region. The results of the present calculation coincide with those theoretical approaches, predicting g-factors nearly constant in $^{164}$Dy. It should be pointed out that, in the context of the present study, we cannot justify the subtle changes in the value of the g-factors by the presence of neutrons in the intruder orbitals such as $i_{13/2}$, which have been explicitly excluded from our valence space.

\begin{table}
\begin{tabular}{c|cc|cc|c|c|c|c}\hline \hline
 & \multicolumn{8}{c}{B(M1) $[\mu_N]^2 \times 10^{-2}$} \\
$J_{i,band}^{\pi} \rightarrow J_{f,band}^{\pi}$ & \multicolumn{2}{c}{$^{160}$Dy} & \multicolumn{2}{c}{$^{162}$Dy} & $^{164}$Dy & $^{166}$Dy & $^{168}$Dy & $^{170}$Dy \\
                                        &  Exp.  & Theo.   &   Exp.      & Theo.  & Theo.   & Theo.   & Theo.   & Theo.  \\ \hline
$2^+_{gsb}  \rightarrow  2^+_{\gamma}$ &0.01$\pm$0.001&0.13&0.001$\pm$0.0001&0.13 &  0.10   &  *      &  *      &  0.06  \\
$2^+_{gsb}  \rightarrow  3^+_{\gamma}$  &        &  0.26   &             & 0.27   &  0.19   &  0.06   &  0.05   &  0.06  \\
$4^+_{gsb}  \rightarrow  3^+_{\gamma}$  &        &  0.12   &             & 0.12   &  0.09   &  0.07   &  0.07   &  0.08  \\
$4^+_{gsb}  \rightarrow  4^+_{\gamma}$  &        &  0.38   &             & 0.39   &  0.37   &  0.19   &  0.19   &  *     \\
$4^+_{gsb}  \rightarrow  5^+_{\gamma}$  &        &  0.56   &             & 0.57   &  0.44   &  0.04   &  0.04   &  0.10  \\
$6^+_{gsb}  \rightarrow  5^+_{\gamma}$  &        &  0.35   &             & 0.35   &  0.28   &  0.03   &  0.04   &  0.41  \\
$6^+_{gsb}  \rightarrow  7^+_{\gamma}$  &        &  0.93   &             & 0.93   &  0.76   &  0.08   &  0.01   &  0.08  \\
$2^+_{gsb}  \rightarrow  2^+_{\beta}$   &0.39$\pm$0.13&0.15&             & 0.03   &  0.06   &  0.01   &  0.01   &  0.02  \\
$4^+_{gsb}  \rightarrow  4^+_{\beta}$   &        &  0.07   &             & 0.01   &  0.24   &  0.10   &  0.07   &  0.07  \\
$6^+_{gsb}  \rightarrow  6^+_{\beta}$   &        &  0.32   &             & 0.01   &  0.55   &  *      &  0.25   &  0.22  \\
$8^+_{gsb}  \rightarrow  8^+_{\beta}$   &        &  0.03   &             & 0.05   &  1.28   &  0.12   &  0.07   &  0.56  \\
$2^+_{\beta}  \rightarrow  2^+_{\gamma}$&        &  0.15   &             & 0.45   &  *      &  0.03   &  0.04   &  0.61  \\
$2^+_{\beta}  \rightarrow  3^+_{\gamma}$&        &  0.77   &             & 0.11   &  0.06   &  0.24   &  0.28   &  0.07  \\
$4^+_{\beta}  \rightarrow  3^+_{\gamma}$&        &  1.02   &             & 0.15   &  0.05   &  0.30   &  0.41   &  2.18  \\
$4^+_{\beta}  \rightarrow  4^+_{\gamma}$&        &  0.31   &             & 0.11   &  *      &  0.11   &  0.11   &  0.46  \\
$4^+_{\beta}  \rightarrow  5^+_{\gamma}$&        &  2.99   &             & 0.43   &  0.44   &  0.05   &  0.03   &  2.36  \\
$6^+_{\beta}  \rightarrow  5^+_{\gamma}$&        &  1.02   &             & 1.85   &  0.31   &  0.34   &  0.38   &  5.63  \\
$6^+_{\beta}  \rightarrow  6^+_{\gamma}$&        &  0.47   &             & 1.32   &  *      &  *      &  *      &  0.11  \\
$6^+_{\beta}  \rightarrow  7^+_{\gamma}$&        &  1.78   &             & 1.63   &  1.28   &  0.08   &  0.06   &  6.50  \\
\hline \hline
\end{tabular}
\caption{B(M1) values for selected transitions in the low-lying positive-parity spectrum of $^{160-170}$Dy. The left column denotes the initial ($J^\pi_i$) and final angular momentum ($J^\pi_f$), and the columns from second to ninth list the experimental data and the predictions of the pseudo-SU(3) model for B(M1;$J^+_i \rightarrow J^+_f)$, in $[\mu_N]^2 \times 10^{-2}$. The * is for B(M1;$J^+_i \rightarrow J^+_f) < 0.01 \times 10^{-2} [\mu_N]^2$.}
\label{m1}\end{table}

In order to complete the analysis of the magnetic properties and to allow for future comparisons of the M1 strength or the E2/M1 multipole mixing ratio with data, the pseudo-SU(3) predictions for reduced magnetic dipole transition are given in Table \ref{m1}. To evaluate the M1 transition operator between eigenstates of the Hamiltonian (\ref{eq:ham}), the pseudo-SU(3) tensorial expansion of the T1 operator (\ref{mdo}) was employed. 

\section{Summary and conclusions}\label{summary}

An extended version of the pseudo-SU(3) model which includes pseudo-spin 0 and 1 states has been implemented 
to describe the electromagnetic properties of positive parity low-energy states in the rare-earth nuclei $^{160-170}$Dy. By using a systematically parametrized Hamiltonian and the best fit of three parameters for a set of nuclei \cite{Var13}, the model has allowed the study of the B(E2), quadrupole moments, $g$-factors and M1 transitions of the chain of dysprosium isotopes.

Most inter-band B(E2) transitions found in the present calculation are small compared with intra-band, showing that the wave function of the states are almost orthogonal. However, few strengths are one order of magnitude bigger, indicating a greater mixing between ground-state, $\gamma$ and $\beta$ bands. The predictions for $g$-factors agree with experimental results and predictions of other theoretical schemes: a constancy in the value of $g(2_1)$ for mid-shell nuclei. In reference to the ratio $r=g(2^+_\gamma)/g(2^+_{gsb})$, the pseudo-SU(3) model predicts, as does the Projected Shell Model, a transition towards triaxiality when we move from $^{160}$Dy to $^{170}$Dy nucleus.

In the present contribution, the work has been focussed on dysprosium isotopes. A logical continuation is to apply the model to heavier rare-earth nuclei as Er, Yb or Hf, where the transition towards triaxiality could be more evident. An interesting continuation of the present work is the application of the model to the study of the nature of the $0^+$ excited states and its relation with the $\beta$ excitations. Furthermore, the scheme employed in the present approach can also be used to describe electromagnetic properties in odd mass or odd-odd nuclei.

\section{Acknowledgments}
This work was supported in part by CONACyT (M\'exico). We would like to thank Dr. F. Mason Lambert for the style corrections of the manuscript.

\end{document}